\documentclass[fleqn,usenatbib]{mnras}

\usepackage{newtxtext,newtxmath}
\usepackage[T1]{fontenc}
\usepackage{ae,aecompl}
\usepackage{float}
\usepackage{afterpage}
\usepackage[toc,page]{appendix}

\usepackage{graphicx}	% Including figure files
\usepackage{amsmath}	% Advanced maths commands
\usepackage{amssymb}	% Extra maths symbols
\usepackage{tabularx}
\usepackage{caption}
\usepackage{ragged2e}
\usepackage[dvipsnames]{xcolor}

\title[X-Ray Spectral Analysis of 1H 0323+342]{The origin of X-ray emission in the gamma-ray emitting narrow-line Seyfert 1 1H 0323+342}

\author[Mundo et al.]{Sergio A. Mundo,$^{1}$\thanks{E-mail: smundo@astro.umd.edu}
Erin Kara,$^{2, 1}$
Edward M. Cackett,$^{3}$
A.C. Fabian,$^{4}$
\newauthor
J. Jiang,$^{5, 4}$
R. F. Mushotzky,$^{1}$
M. L. Parker,$^{6}$
C. Pinto,$^{7}$
C. S. Reynolds,$^{4}$
A. Zoghbi$^{8}$
\\
$^{1}$Department of Astronomy, University of Maryland, College Park, MD 20742, USA\\
$^{2}$MIT Kavli Institute for Astrophysics and Space Research, Cambridge, MA 02139, USA\\
$^{3}$Department of Physics \& Astronomy, Wayne State University, 666 W. Hancock St, Detroit, MI 48201, USA\\
$^{4}$Institute of Astronomy, University of Cambridge, Madingley Road, Cambridge CB3 0HA, UK\\
$^{5}$Tsinghua Center for Astrophysics, Tsinghua University, Beijing 100084, China\\
$^{6}$European Space Agency (ESA), European Space Astronomy Centre (ESAC), E-28691 Villanueva de la Ca\~{n}ada, Madrid, Spain\\
$^{7}$ESTEC/ESA, Keplerlaan 1, 2201AZ Noordwijk, The Netherlands\\
$^{8}$Department of Astronomy, University of Michigan, Ann Arbor, MI 48109-1107, USA 
}

\date{Accepted 2020 June 5. Received 2020 April 24; in original form 2019 August 29}

\pubyear{2020}

\begin{document}
\label{firstpage}
\pagerange{\pageref{firstpage}--\pageref{lastpage}}
\maketitle

\begin{abstract}
We present the results of X-ray spectral and timing analyses of the closest gamma-ray emitting narrow-line Seyfert 1 ($\gamma$-NLS1) galaxy, 1H 0323+342. We use observations from a recent, simultaneous \textit{XMM-Newton}/\textit{NuSTAR} campaign. As in radio-quiet NLS1s, the spectrum reveals a soft excess at low energies ($\lesssim2$ keV) and reflection features such as a broad iron K emission line. We also find evidence of a hard excess at energies above $\sim35$ keV that is likely a consequence of jet emission. Our analysis shows that relativistic reflection is statistically required, and using a combination of models that includes the reflection model \texttt{relxill} for the broadband spectrum, we find an inclination of $i=63^{+7}_{-5}$ degrees, which is in tension with much lower values inferred by superluminal motion in radio observations. We also find a flat ($q=2.2\pm0.3$) emissivity profile, implying that there is more reflected flux than usual being emitted from the outer regions of the disk, which in turn suggests a deviation from the thin disk model assumption. We discuss possible reasons for this, such as reflection off of a thick accretion disk geometry.          
\end{abstract}


%%%%%%%%%%%%%%%%%%%%%%%%%%%%%%%%%%%%%%%%%%%%%%%%%%

%%%%%%%%%%%%%%%%% BODY OF PAPER %%%%%%%%%%%%%%%%%%

\section{Introduction}

Narrow-line Seyfert 1 (NLS1) galaxies are a type of active galactic nucleus (AGN) characterized by their unique optical spectral features, such as broad Balmer emission lines with low widths (FWHM$_{\mathrm{H}\beta}<2000$ km s$^{-1}$), weak [OIII] emission ([OIII]/H$_{\beta}$ flux $<$ 3), and strong FeII emission \citep[]{1985ApJ...297..166O}. Studies have suggested that NLS1s host supermassive black holes on the lower-mass end ($\sim 10^{6}-10^{8} \mathrm{M}\textsubscript{\(\odot\)}$, \citep{2004ApJ...606L..41G,2006AJ....132..321D}) that accrete near or above the Eddington limit \citep[e.g.,][]{2000ApJ...542..161P}. As a result, it is believed that NLS1s form a set of younger AGN that have yet to transform into more luminous quasars. 

In addition to their optical properties, NLS1s are bright in the X-ray band and exhibit complex X-ray spectral features, such as an excess in soft X-ray emission and reflection features in the hard X-rays. As with other accretion systems around black holes (e.g. black hole X-ray binaries), the primary form of the X-ray emission is a continuum well-modeled by a power law that results from the inverse Compton scattering of seed disk photons by a hot plasma of electrons, or a corona, that lies above the disk in the vicinity of the black hole. Some of this continuum emission illuminates the accretion disk, and the upscattered photons end up either Compton scattering off of electrons in the disk, or are reprocessed through fluorescence (see \citealt{2003PhR...377..389R} for a review). These reflection mechanisms have also been featured in the spectra of NLS1s in the form of an iron emission line between 6-7 keV and a Compton reflection ``hump" that peaks between around 20 and 30 keV, implying that the reflection is occurring off of an ionized disk (for reflection features in NLS1s, see e.g. \citealt{2014MNRAS.440.2347M,2017MNRAS.468.3489K}). The Fe emission line in these spectra is usually broadened and skewed towards lower energies due to line-of-sight Doppler boosting and the gravitational potential of the black hole, respectively \citep{1989MNRAS.238..729F,2003PhR...377..389R}. Which effect is most dominant depends on the inclination of the disk relative to our line of sight; therefore, from reflection processes alone, we can arrive at an estimate for the disk inclination of a NLS1.

The origin of the soft excess is still disputed. It can be modeled as blackbody thermal emission, but this is not physical because the resulting blackbody temperature is simply too high to be emission from the disk. A way around this is to consider a ``warm" Comptonization region that is optically thicker than the corona, which yields a temperature of around 0.1-0.2 keV that is constant across a wide range of black hole masses and accretion rates \citep[e.g.,][]{2004MNRAS.349L...7G}. Other ideas that have been put forth place atomic processes like reflection or absorption as the culprits. One example of the former is that coronal illumination of the disk could result in the fluorescence of lines at lower energies that end up being blurred due to gravitational effects \citep{2006MNRAS.365.1067C,2009Natur.459..540F,2013MNRAS.428.2901W}. The soft excess could also be described by a disk with a high electron density. With a higher density, bremsstrahlung would have a higher contribution to the spectrum, in the form of an increased temperature at the surface of the illuminated disk that results from free-free absorption \citep{2016MNRAS.462..751G,2018MNRAS.477.3711J,2018MNRAS.479..615M,2019MNRAS.489.3436J}. This, in turn, may cause blurred reflection at low energies to look more like a blackbody spectrum. If a higher density disk is not taken into account, this could result in a perceived excess at lower energies, especially in AGN with smaller supermassive black holes ($M \ll 10^{9} \rm M_{\odot}$). 

The spectral features of NLS1s have also been described by a series of alternative, absorption-based partial covering models. In this family of models, the X-ray  variability is not intrinsic to the source, but is rather caused by a varying partial covering fraction of clouds that possibly result from disk instabilities or radiation-driven outflows. \citealt{2004MNRAS.353.1064G} and \citealt{2004PASJ...56L...9T} described spectral changes in the NLS1 1H 0707-495 with neutral single and double layer absorbers, respectively, with the latter assuming that the covering fraction changed with the clouds' orbital motion. In addition, \citealt{2012PASJ...64..140M} and \citealt{2014PASJ...66..122M} were able to explain the broad line feature in MCG-6-30-15 and 1H 0707-495 with ionized partial covering models, with \citealt{2014PASJ...66..122M} suggesting that the clouds are produced by funnel-shaped disk winds. However, several studies have shown that most accreting objects have a linear RMS-flux relation, which suggests that the underlying X-ray variability processes are multiplicative in nature and are therefore intrinsic to the source \citep[e.g.,][]{2019MNRAS.485..260A,2005MNRAS.359..345U}. This is at odds with the inherent features of partial covering models, which would show shot noise, additive variability. Due to these characteristics and the mysterious nature of some of these observations, NLS1s, along with other Seyferts, are not only ideal candidates for studying X-ray emission processes and their origins, but also offer interesting possibilities in the realm of X-ray astronomy pertaining to an AGN's central region.

Roughly 10\% of AGN exhibit collimated, relativistic jets that emit in the radio band via synchrotron radiation \citep[e.g.,][]{1984RvMP...56..255B}. Another unsolved problem in the physics of AGN processes is exactly where and how these jets are launched, and as a result, the connection between the disk, corona, and the jet's driving mechanism is currently poorly understood. In any case, ``radio-loud" (RL) AGN with the X-ray spectral features discussed earlier would provide the most promising environment in which to study the interaction between these three and other components. Most NLS1s, however, are ``radio-quiet" (RQ), meaning that their jets are not nearly as powerful as those from blazars or other RL AGN. Therefore, looking for the long-sought disk-corona-jet connection in these AGN proves to be quite difficult.

In recent years, a new class of NLS1s has been discovered by the \textit{Fermi Gamma-Ray Telescope}. These are gamma-ray emitting, RL NLS1s ($\gamma$-NLS1), which have features seen in both NLS1s (discussed previously) and blazars, such as flat radio spectra, double-hump spectral energy distribution (SED) \citep[e.g.,][]{2009ApJ...707L.142A}, and occasionally superluminal motion \citep[e.g.,][]{2013MNRAS.436..191D}. The presence of gamma-ray emission in these galaxies is evidence for a jet, given that in those blazars classified as flat spectrum radio quasars, photons from outside a jet can be inverse Compton scattered to hard X-rays or $\gamma$-rays by the relativistic particles in the jet; this is referred to as external Compton (EC) \citep[e.g.,][]{2000ApJ...545..107B,2009MNRAS.396L.105G}. Studies of the X-ray spectra of these $\gamma$-NLS1s have already shown not only a soft excess, but also a ``hard excess'' above a few keV that requires a much harder power law component compared to the ones usually found in AGN \citep{2014MNRAS.440..106B}, possibly representing the jet emission. Therefore, these $\gamma$-NLS1s provide us with an unusual laboratory for studying properties from both the jet and the thermal emission from the corona simultaneously, and can give us insight into the dominant mechanism that produces the X-ray emission, while at the same time shedding light on how extragalactic jets are formed.

The closest of these exotic AGN is 1H 0323+342 ($\alpha$: 03 24 41.16, $\delta$: +34 10 45.8), with a redshift of $z = 0.06$ \citep{2007ApJ...658L..13Z}. Radio images of this source exhibit superluminal motion (1-7 times the speed of light), which in turn implies the presence of a relativistic jet at an angle $i = 4-13$ deg from the line of sight \citep{2016RAA....16..176F}. 1H 0323+342 also has a double-hump SED characteristic of $\gamma$-NLS1s that peaks in the radio band and gamma-rays, implying synchrotron emission and synchrotron self-Compton (or EC) mechanisms that would result from a jet.

Exactly where the X-ray emission from 1H 0323+342 comes from is still unknown. Currently, two main explanations have been put forth: it could be the result of interactions between the disk and the corona, as in RQ NLS1s (power-law continuum, reflection features, etc.; see \cite{2014ApJ...789..143P}, \citet{2019ApJ...872..169P}), or it could simply be a continuation of the double-hump SED that fills the gap between the radio and gamma-ray emission from the jet, implying that X-ray emission would be included as a direct consequence of interactions with the jet (for a detailed multi-wavelength spectral analysis for this source, see \citealt{2018MNRAS.475..404K}).

1H 0323+342 has been observed by \textit{Swift} and \textit{Suzaku}, and the spectra show properties seen in RQ NLS1s, such as a soft excess and a potential broad Fe K$\alpha$ emission line \citep[e.g.,][]{2013MNRAS.428.2901W,2019ApJ...872..169P}. Archival data has shown that there is a potentially blue-shifted iron line that would require a disk inclination of nearly 90$^\circ$ due to line-of-sight Doppler boosting \citep{2013MNRAS.428.2901W,2015AJ....150...23Y}, which contradicts data from radio observations since the superluminal motion in the latter indicates that the jet is emitted at an angle close to our line of sight, and therefore also suggests that the disk is face-on. However, the broad iron line has never been clearly detected due to low signal-to-noise.

\begin{table}
	\centering
	\caption{\textit{XMM-Newton} and \textit{NuSTAR} observations used in our analysis. Shown are the detector name, observation ID, observation start dates, the duration of the observations, and the effective exposure times (for \textit{XMM}, after excluding epochs of flaring particle background). All \textit{XMM} observations were made in Large Window mode.}
	\label{tab:obs}
	\begin{tabularx}{\columnwidth}{XXXXX} % four columns, alignment for each
		\hline
		Detector & Obs. ID & Start Date & Duration (s) & Effective Exposure (s)\\
		\hline
		EPIC-pn & 0764670101 & 2015/08/23 & 80900 & 62000\\
		 & 0823780201 & 2018/08/14 & 53066 & 47600\\
		 & 0823780301 & 2018/08/18 & 48212 & 43800\\
		 & 0823780401 & 2018/08/20 & 47975 & 45461\\
		 & 0823780501 & 2018/08/24 & 48515 & 46002\\
		 & 0823780601 & 2018/09/05 & 50818 & 44700\\
		 & 0823780701 & 2018/09/09 & 49468 & 46951\\
		FPMA/ & 60061360002 & 2014/03/15 & 108880 & 101633\\
		FPMB & 60402003002 & 2018/08/14 & 38937 & 36390\\
		 & 60402003004 & 2018/08/18 & 31711 & 29732\\
		 & 60402003006 & 2018/08/20 & 28137 & 26401\\
		 & 60402003008 & 2018/08/24 & 27276 & 25565\\
		 & 60402003010 & 2018/09/05 & 32516 & 30422\\
		 & 60402003012 & 2018/09/09 & 29921 & 27795\\
		\hline
	\end{tabularx}
\end{table}

\begin{figure*}
	\includegraphics[width=\textwidth]{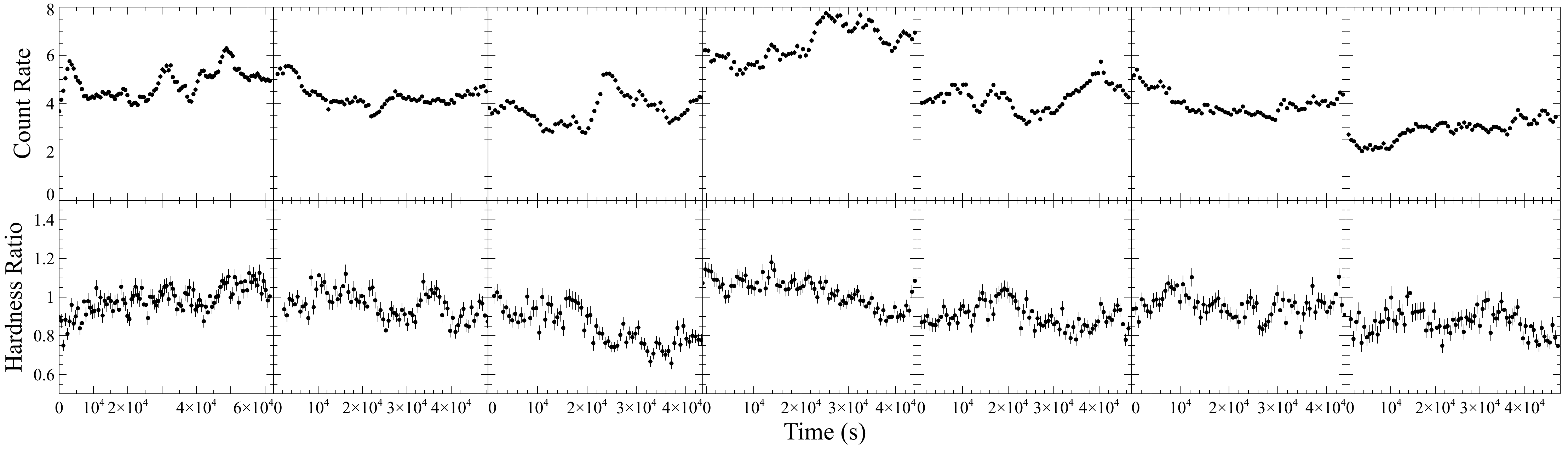}
    \caption{\textit{Top panel}: EPIC-pn broadband lightcurves from 0.3-10 keV in 600 s bins. Light curves were generated with circular extraction regions of 40 arcsec radii. The first observation is archival and is longer than the rest by about 30 ks. \textit{Bottom panel}: Hardness ratio (1-4 keV to 0.3-1 keV) of the EPIC-pn observations. For the spectral analysis, the observations were co-added as the hardness ratio remains fairly constant throughout the campaign.}
    \label{fig:LC}
\end{figure*}

In this paper, we present an X-ray spectral and timing analysis with the first such data set from \textit{XMM-Newton} and \textit{NuSTAR} for this source. We aim to find the origin of the X-ray emission in 1H 0323+342, while at the same time obtaining an estimate of the disk inclination purely from the X-ray spectra. We describe the observations and data reduction in Section \ref{sec:obs}, present our spectral and timing analyses and their results in Section \ref{sec:res}, and discuss these results in Section \ref{sec:disc}.

\section{Observations and Data Reduction}
\label{sec:obs}

To study this source, we use a set of six simultaneous observations made by both \textit{XMM-Newton} and \textit{NuSTAR} (twelve total, PI: Kara; see Table \ref{tab:obs}), as well as an archival observation that was made in 2015 (ID 0764670101; PI: D'Ammando).

\subsection{\textit{XMM-Newton}}
\label{sec:xmm} % used for referring to this section from elsewhere

The \textit{XMM-Newton} data were reduced with the \textit{XMM-Newton} Science Analysis System (SAS v.16.1.0) and the current calibration files available. We focus on the data from the EPIC-pn camera here, which was taken in Large Window mode. All observations were checked for flaring particle background, and we set \textsc{pattern} $\leq 4$ to choose single and double events and \textsc{flag} $== 0$ to get the best quality data. The data was also checked for pile-up with the SAS epatplot task, but there was no pile-up present in any of the observations. We extracted the source spectra from circular regions with radii of 40$^{\prime\prime}$ that were centered on the J2000 coordinates of 1H0323+342. The background spectra were obtained from circular background regions that were offset from the source and were made as large as possible, with radii of 70$^{\prime\prime}$. 

Background-subtracted light curves were also extracted with the \textsc{epiclccorr} tool and were binned with 10 second bins (see Fig. \ref{fig:LC}). Although the flux varies by about a factor of 4, there was no significant change in the spectrum between observations, and the hardness ratio between the hard and soft bands remains relatively constant across each observation. We therefore decided to combine them to form one spectrum. This co-added spectrum was rebinned with the \texttt{grppha} tool to ensure that we would have a minimum of 25 counts per bin.

\subsection{\textit{NuSTAR}}

We extracted the \textit{NuSTAR} data using HEASOFT v.6.22.1 and the standard \texttt{nupipeline} task, again from 40$^{\prime\prime}$ circular regions for both source and background spectra, for each focal plane module (FPMA/FPMB). The spectra were also co-added with the archival observation and binned with \texttt{grppha} to have at least 25 counts per bin. The source spectrum is above the background spectrum up to $\sim50$ keV, so we show the relevant plots up to this energy.

\section{Results}
\label{sec:res}

\subsection{RMS-Flux Relation}

In the past, it has been shown that most accreting sources have a linear relationship between absolute RMS and flux. Previous works \citep[e.g.,][]{2019MNRAS.485..260A,2005MNRAS.359..345U,2001MNRAS.323L..26U} have shown that this relation implies a variability process that is intrinsic to the source and that is likely a multiplicative process, building up from propagating mass accretion rate fluctuations occurring in the disk. Since the best model to describe this is one where accretion disk fluctuations are the source of the variability, finding such a relation in 1H 0323+342 would suggest that disk-corona and reflection models, where the X-rays are assumed to be intrinsic variations originating from the innermost regions of the disk, would be best to describe the data.

\begin{figure}
	\includegraphics[width=\columnwidth]{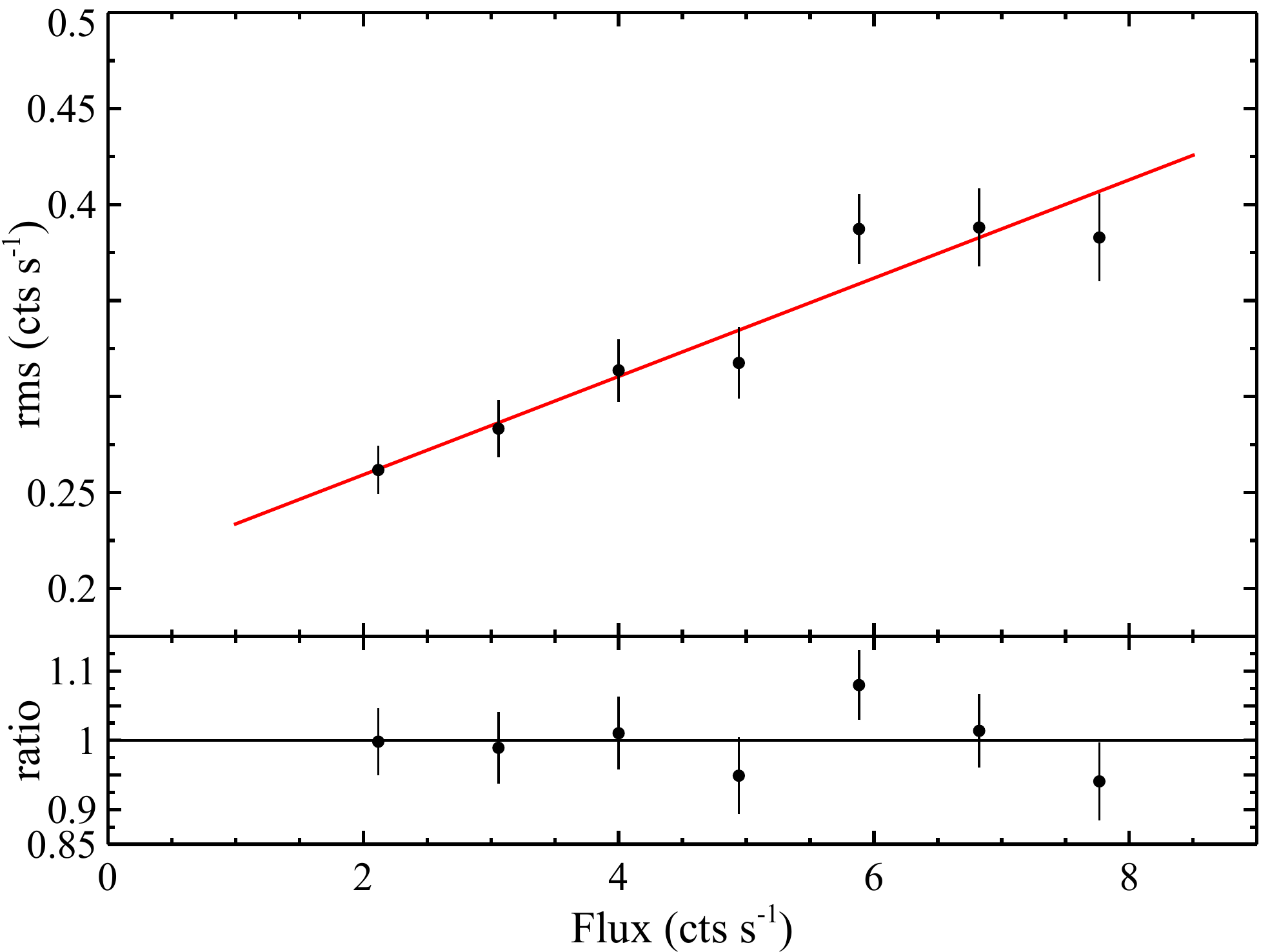}
    \caption{Relationship between absolute rms and flux. On average, the absolute rms increases linearly with flux, suggesting that variations are intrinsic to the source.}
    \label{fig:rmsflux}
\end{figure}

We begin a calculation of the rms-flux relationship in the time domain by binning our \textit{XMM} lightcurves in 1000 s bins. At the same time, we compute the excess variance as a function of time with the same binning, effectively producing a `variance light curve'. We take the square root of the excess variance to obtain the rms amplitude, and sort these values by count rate. We then bin the rms amplitudes by flux such that we have $\sim50$ points in each bin. Errors on the rms amplitude are calculated as in \cite{2003MNRAS.345.1271V}. We show in Fig. \ref{fig:rmsflux} that we indeed obtain a linear rms-flux relation, with a linear fit of $\chi^{2} = 4.61$ for 5 degrees of freedom. This therefore supports the use of reflection models for this data and implies that the X-ray variations are intrinsic.

\subsection{Spectral Features}

We fit the \textit{XMM-Newton} and \textit{NuSTAR} spectra simultaneously using X\textsc{spec} v12.9.1p \citep{1996ASPC..101...17A}. We included an overall multiplicative constant in each of our models ($C_{\rm A}$ for FPMA and $C_{\rm B}$ for FPMB) to take into account differences in absolute flux calibration between each detector. We used the {\tt{tbabs}} model \citep{2000ApJ...542..914W} and cross-sections from \citet{1996ApJ...465..487V} to model the galactic absorption, setting $N_{\mathrm{H,Gal}} = 1.26\times10^{21} \mathrm{cm}^{-2}$ \citep{2005A&A...440..775K}. In all of our fits, we also check for potentially different spectral shapes between \textit{XMM} and \textit{NuSTAR}, but we find that the spectral index $\Gamma$ only differs by $\sim1\%$ and does not affect our results, so we fit with the same photon index. All spectra are plotted in the source rest frame energies.

\subsubsection{Reflection Features above 2 keV}

\begin{figure}
	\includegraphics[width=\columnwidth]{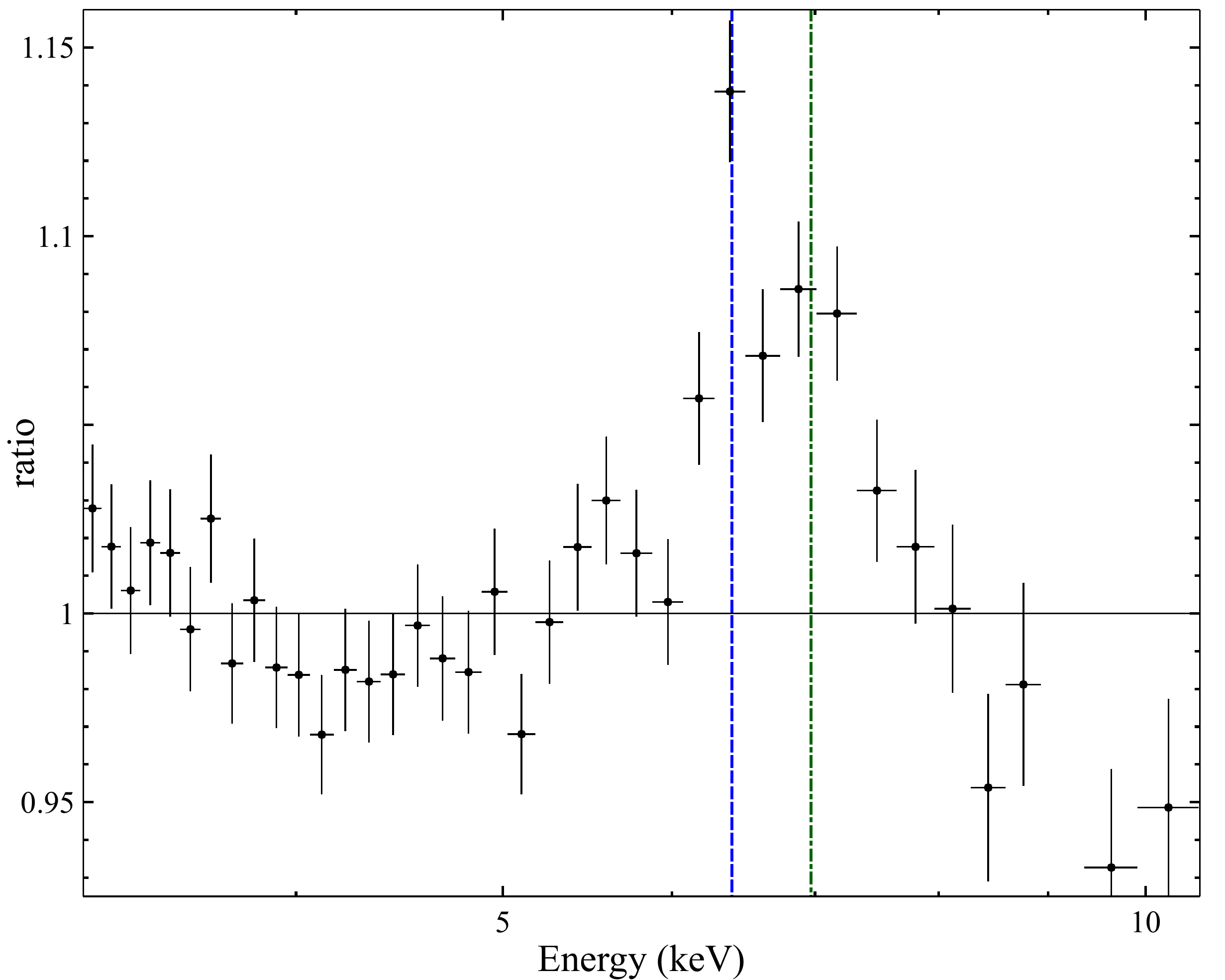}
    \caption{The ratio of EPIC-pn to a simple power law; energies are in the rest frame of the source. The blue line shows the position of neutral iron at 6.4 keV, the green line that of hydrogen-like Fe XXVI at 6.97 keV. We find an equivalent width of $175\pm40$ eV for the broad iron line.}
    \label{fig:Fe}
\end{figure}

We begin by searching for the broad iron line, which has only been marginally detected in previous observations \citep[e.g.,][]{2019ApJ...872..169P,2014ApJ...789..143P,2015AJ....150...23Y,2013MNRAS.428.2901W}. To show the highest resolution version of the line and to show we are making a significant detection, we fit the 3-10 keV spectra with a power law (see Figure \ref{fig:Fe}) and compare this to a power law plus Gaussian. The addition of the Gaussian improves the fit by $\Delta\chi^{2} = 130$ for 3 additional parameters, and therefore a line is preferred at >99.99\% confidence. We find a line at $6.6\pm0.1$ keV in the rest frame of the source that is broad with respect to the spectral resolution of the instruments, with $\sigma = 0.62^{+0.13}_{-0.12}$ keV. This is consistent with Fe K$\alpha$ emission. We also calculate an equivalent width of around $175\pm40$ eV for \textit{XMM}, which is stronger than in archival, lower signal-to-noise observations \citep[e.g.,][]{2018MNRAS.475..404K} and again supports our claim of a broad line.

Figure \ref{fig:Fe} also shows a relatively pronounced peak between 6 and 7 keV, at the 6.4 keV of neutral iron fluorescence. We therefore include a narrower line fixed at 6.4 keV, in addition to the aforementioned Gaussian, and fix its width to 100 eV. This improves the fit by $\Delta\chi^{2} = 10$ for one additional parameter, at a significance of $>99.8$\%.  This suggests that, in addition to the broad iron line, there may also be a narrow component from a distant reflector that contributes to the spectrum.

\begin{figure}
	\includegraphics[width=\columnwidth]{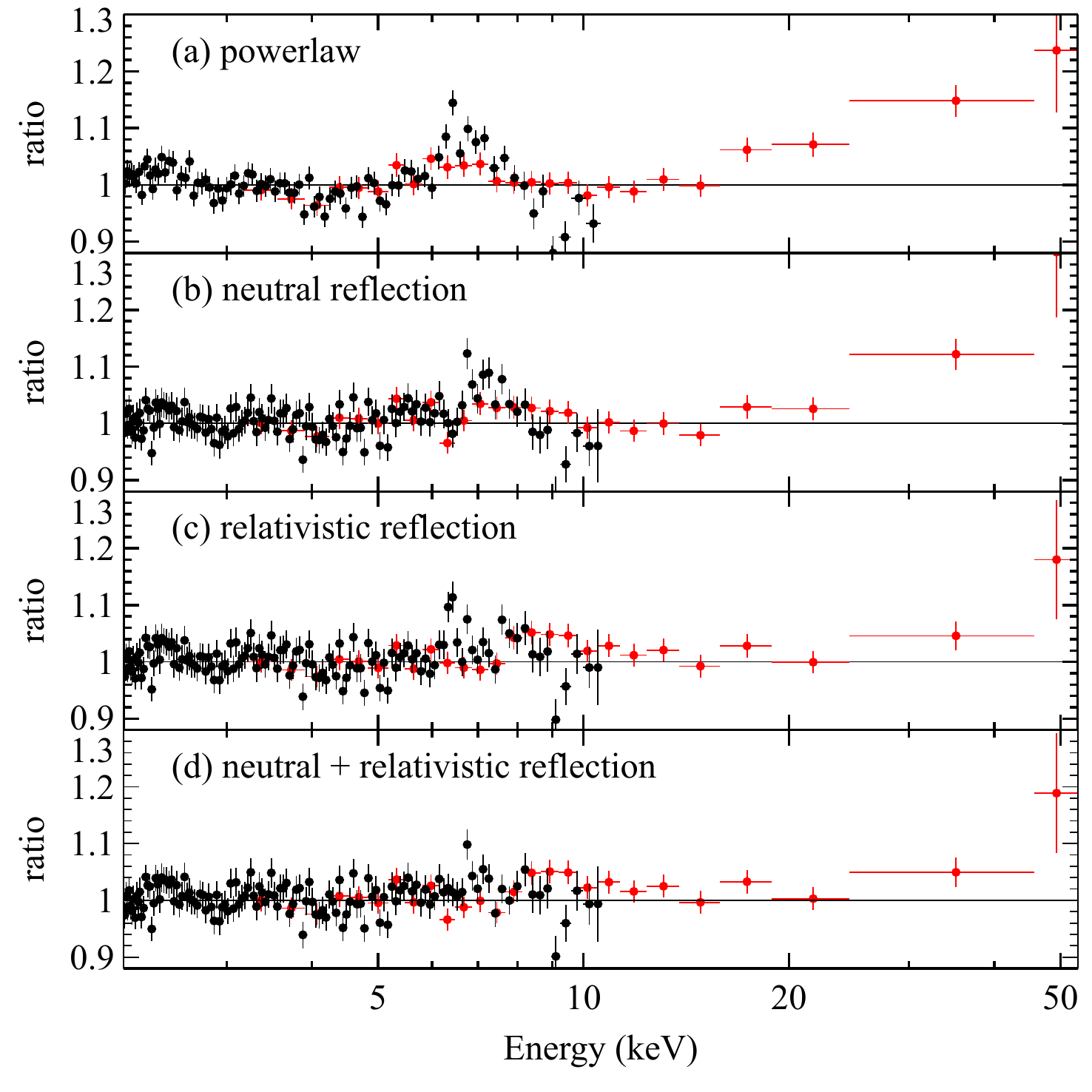}
    \caption{The ratio of EPIC-pn and \textit{NuSTAR} spectra to different models. (a) The ratio to a power law. (b) Ratio to neutral reflection model. (c) Ratio to relativistic reflection model. (d) Ratio to relativistic reflection model with an additional neutral, distant reflector.}
    \label{fig:resids}
\end{figure}

A power law fit to the 2-79 keV spectra of \textit{XMM-Newton} and \textit{NuSTAR} shows not only a broad iron line peaking at 6-7 keV, but also a Compton reflection hump above 10 keV, indicating reflection off of an ionized disk (see Figure \ref{fig:resids}). This power law yields a reduced chi-squared $\chi^{2}/\rm d.o.f = \chi_{\nu}^{2} = 2552/2385 = 1.07$. Given the nature of the spectra, it is clear that reflection and possibly relativistic broadening need to be taken into account, so we fit the 2-79 keV data with different flavors of the reflection model \texttt{relxill} \citep{2016A&A...590A..76D}. Following up on the narrow component at 6.4 keV discussed earlier, we start with a neutral, non-relativistic reflection model \texttt{xillver}, with the ionization parameter log $\xi$ fixed at 0. This gives a fit with $\chi_{\nu}^{2} = 1.01$, improving from the power law fit by $\Delta\chi^{2} = 137$ for 4 additional free parameters, with a significance of $>$ 99.99\% evaluated using the $F$-test. The residuals between 6-7 keV, along with the fact that we detect a fairly broad line, suggest we still need to account for relativistic smearing. 

We use the relativistic model \texttt{relxill} and set the inner accretion disk radius $R_{\rm in}$ to the innermost stable circular orbit. We start by fixing the emissivity index to the Newtonian value $q = 3$ and find a good fit, with $\chi_{\nu}^{2} = 2385/2380 = 1.002$. We also attempt using a broken power law emissivity, but this yields a break radius $R_{\rm break}$ that is below the innermost stable circular orbit, which is not physical. We therefore keep an unbroken emissivity for the rest of the paper. We also cannot constrain the spin $a$. Since this is a jetted source, it may involve a high-spin object, so we use this physical motivation to fix the spin to the maximum value of 0.998.

We combine the models in the second and third panels of Figure \ref{fig:resids} to check for the possibility of an additional neutral, distant reflector. All relevant parameters in {\tt{xillver}} are tied to the parameters of {\tt{relxill}}, except for the normalization. This improves the fit by $\Delta\chi^{2} = 31$ for one additional free parameter, at a significance of $>$ 99.99\%. This fit finds an inclination $i=59^{+4}_{-5}$ deg, which is in tension with the low inclination obtained from radio observations. Within 90\% confidence, this fit yields very similar parameters as the ones with just {\tt{relxill}}, and we adopt this version for later analyses of the broadband spectrum. We report the best-fit parameters in Table \ref{tab:table3}. 

Our fits also seem to show consistent negative residuals at $\sim 9$ keV (see Figures \ref{fig:Fe} and \ref{fig:resids}). It is important to note that this is most likely not physical. When we plot the background spectrum, we see that this is probably a background feature, namely the EPIC-pn Cu-K$\alpha$ complex at around those energies. Regardless of our source region size or background region placement (keeping the latter within the inner 4 CCDs), the feature remains. The residuals are therefore likely the result of over-subtraction of the EPIC-pn background features. We verify this by plotting the spectrum without the background subtracted, and find that the feature is not present (see Appendix \ref{appendix:A}). We also do not see the negative residuals in any of the individual EPIC-MOS spectra.

\begin{table}
    %\large
    %\fontsize{12}{12}\selectfont
    \setlength{\tabcolsep}{6pt}
    \renewcommand{\arraystretch}{1.5}
    \centering
%    \caption{Best-fit parameters}
%    \label{tab:table2}
    \resizebox{\columnwidth}{!}{\begin{tabular}{rccc} 
        \hline \hline
        Parameter & \texttt{xillver + relxill} & \texttt{diskbb + xillver + relxill + po} \\
        \hline
        $N_{\rm H,Gal} \ (10^{22}$cm$^{-2})$* & 0.126 & 0.126 \\
        
        $N_{\rm H,intrinsic} \ (10^{22}$cm$^{-2})$ & ... & $<0.01$ \\
          
%        $K_{\rm powerlaw}(10^{-5})$ & $50_{-29}^{+40}$ & $98^{+11}_{-13}$\\
        
 %       $K_{\mathrm{Gauss}}(10^{-4})$ & \nodata & ...\\
        
        $K_{\mathrm{\texttt{diskbb}}}$ & ... & $200^{+65}_{-66}$\\
        
        $K_{\rm dist. \ reflection} \ (10^{-5})$ & $1.1^{+0.5}_{-0.4}$ & $0.5\pm0.4$\\
        
        $K_{\rm rel. \ reflection} \ (10^{-5})$ & $3.3^{+0.5}_{-0.4}$ & $1.8^{+0.6}_{-0.5}$\\
        
%        $D$ & ... & $0.44^{+0.03}_{-0.04}$\\
        
        $z$* & 0.06 & 0.06\\
        
%        $E_{\rm edge}$ & ... & $0.606^{+0.004}_{-0.003}$\\
        
        $T_{\rm in}$ (keV) & ... & $0.18\pm0.01$\\
         
        $\Gamma$ & $1.91\pm0.01$ & $2.2^{+0.3}_{-0.1}$\\
        
        $\Gamma_{\rm hard}$ & ... & $1.5^{+0.1}_{-0.2}$\\
        
%        $\Gamma_{\rm NuSTAR}$ & $1.82^{+0.03}_{-0.02}$ & $1.83\pm0.02$\\
        
%        $\Gamma_{\rm hard}$ & $1.6^{+0.2}_{-0.5}$ & $1.61\pm0.03$\\
        
%        $\Gamma_{2}$ & ... & $1.55_{-0.09}^{+0.05}$\\
         
        $E_{\rm cut}/kT$ (keV) & $>453$ & $300$\\
        
%        $E_{break}$(keV) & ... & $11_{-3}^{+7}$\\
        
%        Fe line(keV) & ... & ...\\
          
%        $\sigma_{Gauss}$(keV) & ... &\\
        
        $A_{\rm Fe}$ & $3\pm1$ & $4^{+5}_{-2}$\\
          
        log $\xi$ (erg cm s$^{-1}$) & $<2.9$ & $<1$\\
        
%        log $n_{e}$(cm^{-3}) & ... & $17.8^{+0.2}_{-0.8}$\\
        
        $\mathcal{R}_{\rm refl}$ & $0.6\pm0.1$ & $1.3^{+0.4}_{-0.7}$\\ 
        
        $i$ (deg) & $59^{+4}_{-5}$ & $63^{+7}_{-5}$\\
        
        $R_{\rm in} \ (\rm ISCO)$* & 1 & 1\\
        
        $R_{\rm out} \ (R_{\rm g})$* & 400 & 400\\
        
        $R_{\rm break} \ (R_{\rm g})$* & $15$ & $15$\\
        
        $q$ & $3$ & $2.2\pm0.3$\\
        
%        $q_{\rm2}$ & $-5^{+2}_{-5}$ & $1.2^{+0.8}_{-1.7}$\\
        
        $a$* & $0.998$ & $0.998$ \\
        
        $C_{\rm A}$ & $1.17\pm0.01$ & $1.17\pm0.01$\\
        
        $C_{\rm B}$ & $1.20\pm0.01$ & $1.20\pm0.01$\\ \hline
        
%        $C_{\rm BAT}$ & ... & $1.3\pm0.2$\\ \hline
        
        $\chi^{2}$/d.o.f. & 2354/2379 & 2693/2679\\
        
        $\chi_{\nu}^{2}$ & 0.99 & 1.01\\ \hline
    \end{tabular}}
    \caption{ 90\% confidence level parameters for the 2-79 keV and 0.5-79 keV ranges, reported with respect to EPIC-pn. $K$'s are normalizations, $C_{\rm A}$ and $C_{\rm B}$ are the calibration constants for FPMA and FPMB, respectively, and $q$ is the emissivity index. $\Gamma_{\rm hard}$ is the photon index of the hard power law, $T_{\rm in}$ the temperature at the inner disk radius, and $E_{\rm cut}$ the cutoff energy in the rest frame of the source. $\mathcal{R}_{\rm refl}$ and $\xi$ are the reflection fraction and ionization parameter of the relativistic reflection component, respectively. $N_{\rm H,Gal}$, $R_{\rm in}$, $z$, $R_{\rm break}$, $a$ and $R_{\rm out}$ were frozen to the corresponding values (parameters with asterisk). In addition, the ionization parameter in the {\tt{xillver}} distant reflector model was frozen to 0.}
    \label{tab:table3}
\end{table}

\subsubsection{Broadband Spectrum}

Fitting a power law to the data in the 2-79 keV range and extrapolating the fit to lower energies also reveals a soft excess below 2 keV. We begin our analysis in the broadband by including a phenomenological disk blackbody component to help fit the spectrum at soft energies. To account for the possibility of intrinsic absorption, we use \texttt{ztbabs}. We also keep the neutral, distant reflector from the previous section.

\begin{figure}
	\includegraphics[width=\columnwidth]{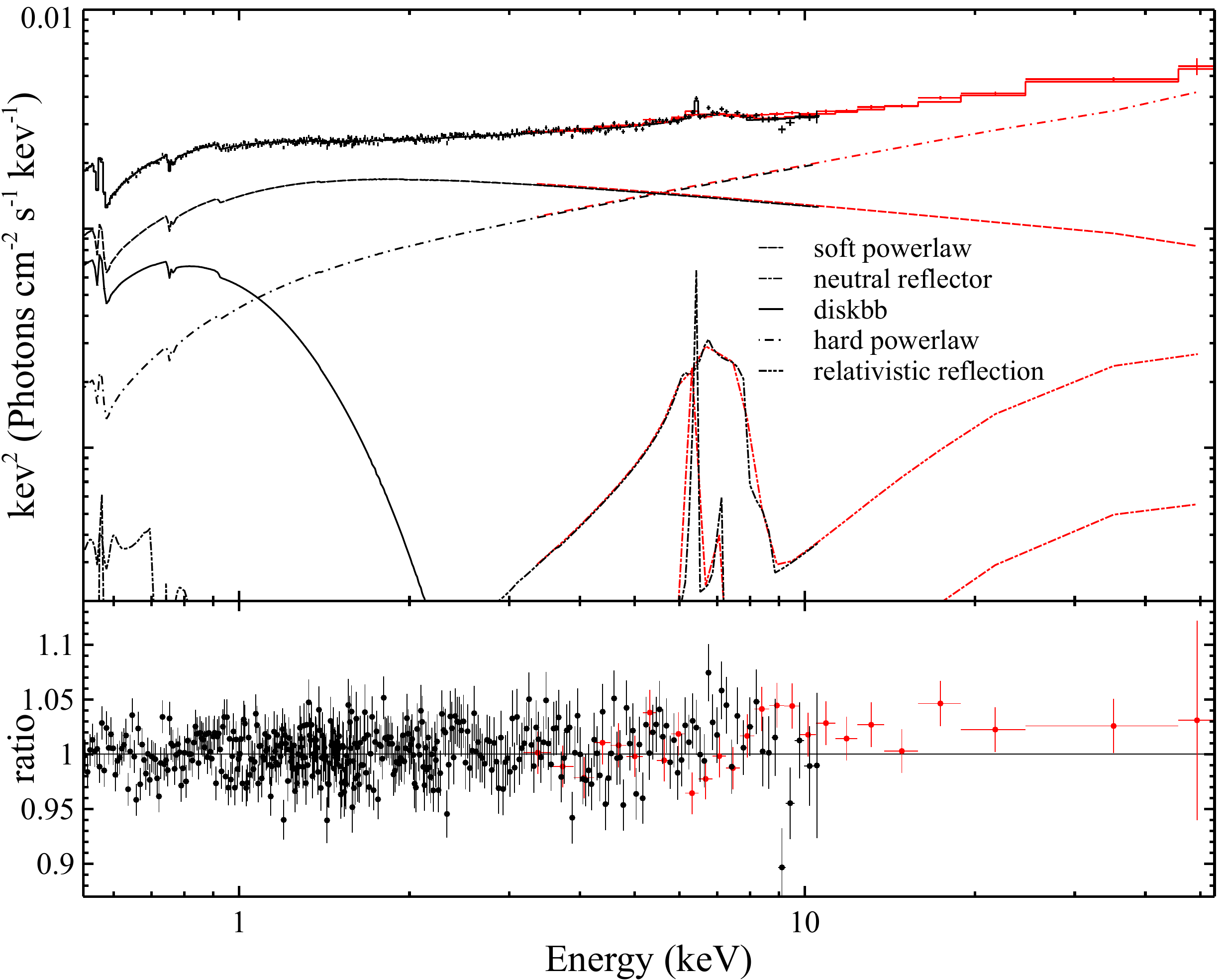}
    \caption{Unfolded spectrum and best-fit model. Model components are also shown.}
    \label{fig:bestfit}
\end{figure}

We first attempt a Newtonian emissivity profile for the \texttt{tbabs*ztbabs*(diskbb + xillver + relxill)} model. This gives a fit with $\chi_{\nu}^{2} = 1.04$. Freeing the emissivity index improves the fit by $\Delta\chi^{2} = 37$ for one additional free parameter at a significance $>$ 99.99\%, with $\chi^{2}/\rm d.o.f. = 2760/2680 = 1.03$. Increasing residuals at energies $\gtrsim 35$ keV (see Figure \ref{fig:resids}) remain, so we include an extra power law to account for a possible contribution of Compton upscattering further out in the jet, similar to \citealt{2019ApJ...872..169P}. An additional hard power law improves the fit by $\Delta\chi^{2} = 46$ for two additional free parameters with a significance $>$ 99.99\%. We find that the fit does not significantly depend on the cutoff energy, so we fix it to 300 keV. This best fit (see Figure \ref{fig:bestfit} and second column of Table \ref{tab:table3}) yields a photon index of $\Gamma = 2.2^{+0.3}_{-0.1}$ for the coronal emission models and  $\Gamma_{\rm hard} = 1.5^{+0.1}_{-0.2}$ for the hard power law, which is consistent with values found in $\gamma$-NLS1s (see e.g. \citealt{2019ApJ...872..169P} and the $\Gamma$ histograms for their double power-law fit, as well as photon index values in \citealt{2020ApJ...896...95O}). It also gives a high inclination of $i=63^{+7}_{-5}$ deg, as before. Moreover, we obtain a relatively flat emissivity index $q=2.2\pm0.3$, suggesting not only that strong general relativistic effects are not required to accurately describe the illumination pattern of the disk, but also that more reflected flux is emitted from the outer regions of the disk. The fact that the observed flux seems to drop slowly at large radii could be an indication that the razor-thin disk assumption is invalid.

We also attempt to model the data with a more physically motivated approach. Following the possibility that the soft excess could arise from reflection off of a high-density disk \citep[e.g.,][]{2016MNRAS.462..751G,2018MNRAS.477.3711J,2018ApJ...855....3T}, we try to fit the soft excess with the extended reflection model \texttt{relxillD}. Given that this source is at a low galactic latitude, it is likely that there is a fair amount of uncertainty in the column density, so we allow \texttt{tbabs} to be a free parameter, and remove \texttt{ztbabs}. We also include an additional hard power law to compare directly with our \texttt{diskbb} fit. This \texttt{tbabs*(xillver + relxillD + po)} model yields an acceptable fit with $\chi^{2}/\rm d.o.f. = 2726/2680$ and an electron density log $n_{e}=17.2^{+0.5}_{-0.1}$, but it requires an iron abundance of $>9$ and a high emissivity index $q = 7^{+1}_{-2}$, which suggests a much higher degree of relativistic smearing than is expected given the equivalent width of our iron line. As in our previous model, it also requires a very high inclination, with $i = 72^{+3}_{-5}$.

For completeness, we also fit the broadband spectrum with an alternative, ionized partial covering model instead of relativistic reflection, replacing {\tt{relxill}} with {\tt{zxipcf*powerlaw}} in our best fit, in order to see if we actually require a broad iron line. We also apply {\tt{zxipcf}} to the disk blackbody component and remove the hard power law component for simplicity. This model, which instead fits the broad line region with a combination of an absorbed power law and a narrow reflection component, gives an acceptable fit with $\chi^{2}/ \rm d.o.f = 2726/2681=1.02$. It requires a covering fraction of $0.56^{+0.07}_{-0.08}$, as well as an outflow velocity of nearly $\sim0.1 c$, which is probably needed in order to account for the blueshift of the iron K line. However, the physical processes represented by this model would manifest themselves in the variability as shot noise, or additive, variability processes, but we instead see a linear RMS-flux relationship representing multiplicative variability processes.

\subsubsection{Disk Inclination}

\begin{figure}
	\includegraphics[width=\columnwidth]{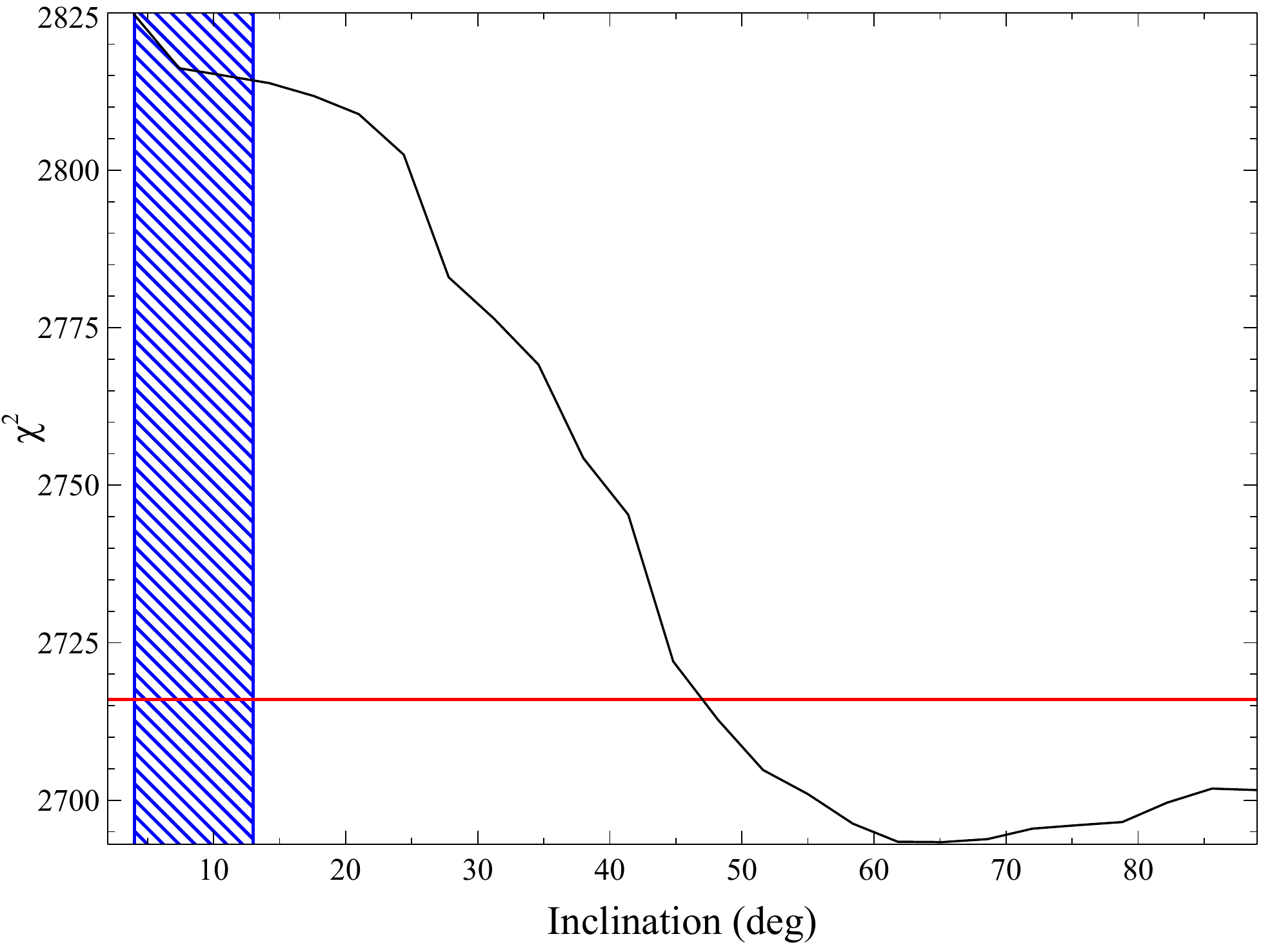}
    \caption{$\chi^{2}$ statistic vs. disk inclination for our best fit. The red horizontal line is the 90\% confidence level. The blue hashed region depicts the range set by radio observations (4-13 deg).}
    \label{fig:Inclorig}
\end{figure}

In our best-fit model, we find an unphysically high disk inclination. We confirm this result by generating the statistic surface for this parameter, which gives comparable values (see Figure \ref{fig:Inclorig}). This is much higher than predicted by superluminal motion in radio observations \citep{2016RAA....16..176F}, which implies a relativistic jet at an inclination $i=4-13$ deg from the line of sight. We therefore attempt to further constrain the inclination angle by placing hard limits at these values. However, we are unable to constrain the inclination this way, as the parameter is pegged at the hard upper limit, so we proceed to fix the inclination to a value of 10 deg. This scenario provides a fit with $\chi_{\nu}^{2} = 1.04$, and we again find that our best fit with an unrestricted inclination is an improvement at a significance of $>99.99\%$ ($\Delta \chi^{2} = 70$ for 1 additional free parameter). Therefore, our data suggests that, although we do observe a broad iron line, the broadening may be caused almost exclusively by line-of-sight Doppler effects that result from a high inclination. However, this is at odds with superluminal motion inferred from radio observations. Alternatively, the geometry is not a simple geometrically thin disk, as assumed by the blurred reflection model.

\subsection{Timing Analysis}

In addition to the spectral analysis described in the previous section, we also performed a timing analysis to compare to timing analyses from RQ NLS1s. X-ray reverberation lags have been used to probe the inner accretion disk region as a means to gain insight into the origin of the X-ray emission in NLS1s \citep[e.g.,][]{2009Natur.459..540F,2014A&ARv..22...72U}. These so called ``soft lags" are caused by a reflection-induced soft excess that lags behind coronal power-law photons due to the short, extra time the latter take to travel from the corona to the inner disk. The detection of a soft lag therefore provides evidence that the soft excess is due to relativistic reflection; on the other hand, a non-detection of a soft lag could suggest that the soft excess is possibly a different component that is not simply reflection.

In the next subsections, we look into whether we can detect soft lags in 1H 0323+342, and analyze the relationship between variations in the soft band and those in the hard band by using Fourier techniques outlined in works such as \citet{1999ApJ...510..874N} and \citet{2014A&ARv..22...72U}.

\subsubsection{Lag-frequency spectrum}

We begin the second part of our timing analysis by using the available 7 light curves to compute time lags between the power law dominated hard band (1-4 keV) and the soft excess (0.3-1 keV) as a function of temporal frequency, or inverse timescale. If the light curve in, say, the soft band, $s$, has $N$ time bins of width $\Delta t$, then its discrete Fourier transform at each Fourier frequency $f_{n} = n/(N\Delta t)$ is
\begin{equation}
    S = \sum_{k=0}^{N-1}s_{k}e^{2\pi ink/N}
\end{equation}
We focus on frequencies higher than $6\times10^{-5} \rm Hz$ to avoid red noise leakage at lower frequencies. We require at least 10 frequencies per bin.

We can rewrite the Fourier transform in a complex polar form as $S = |S|e^{i\phi_{s}}$. Repeating the process for the hard band light curve $h$, we can write the complex conjugate of its Fourier transform as $H^{*} = |H|e^{-i\phi_{h}}$. The product of $S$ and $H^{*}$ gives us the Fourier cross-spectrum between the two bands:
\begin{equation}
    C = |H||S|e^{i(\phi_{s}-\phi_{h})}
\end{equation}
This gives the phase difference between the soft and hard bands, and the average lag between the two bands in each frequency bin is then obtained by taking the phase of the averaged cross-spectrum, or $\phi =$ arg$[\langle C \rangle]$. This can then be manipulated to give the time lag at each frequency bin:
\begin{equation}
    \tau = \frac{\phi}{2\pi f}
\end{equation}

Figure \ref{fig:coh} shows the resulting lag-frequency spectrum between lightcurves in the 0.3-1 keV and 1-4 keV bands that had 10 s bins. To decide our upper limit in frequency, we calculate the frequency range where we expect to see reverberation, given the black hole mass of this AGN, by using the correlation between soft lag frequencies and black hole mass from \citet{2013MNRAS.431.2441D}. \citet{2017MNRAS.464.2565L} obtain a mean mass of $1.9^{+0.4}_{-0.3}\times10^{7} \rm M_{\odot}$ through estimates based on the ionizing 5100\ \AA \ continuum luminosity and the width of the hydrogen broad emission lines. We use this mass in the correlation in \citet{2013MNRAS.431.2441D}, and find that this corresponds to a frequency range $\nu_{\rm lag}=(2.3\pm0.5)\times10^{-4}$ Hz where we can expect to find a soft lag. As a precaution, we extend the frequency range to $1\times10^{-3} \rm Hz$, to account for uncertainties in the mass estimate and the frequency-mass correlation. 

While our spectrum shows a low-frequency hard lag that is consistent with propagating fluctuations in the disk, our spectrum shows no soft lags at shorter timescales, which could be related to a lack of coherence between the soft and hard bands, something we investigate in the next subsection. This non-detection may also in fact support our spectral results by suggesting that reflection off of the accretion disk is not the dominant process causing the soft excess.

\subsubsection{Coherence-frequency spectrum}
In an attempt to further explain the non-detection of soft lags, we proceed to calculate the coherence between the soft and hard light curves. The coherence provides us with a way of measuring how correlated two light curves or signals are. Essentially, it tells us to what extent one light curve can be predicted from the other. The coherence at each frequency is defined as

\begin{equation}
    \gamma^{2} = \frac{|\langle C \rangle|^{2} - n^{2}}{\langle P_{s} \rangle \langle P_{h} \rangle}
\end{equation}
where the $n^2$ term is due to Poisson noise contributing to the square of the cross-spectrum, and the terms in the denominator are noise-subtracted power spectra. For unity coherence, the two light curves would be perfectly coherent, meaning one would be able to predict a light curve from the other through a linear transformation.

\begin{figure}
	\includegraphics[width=\columnwidth]{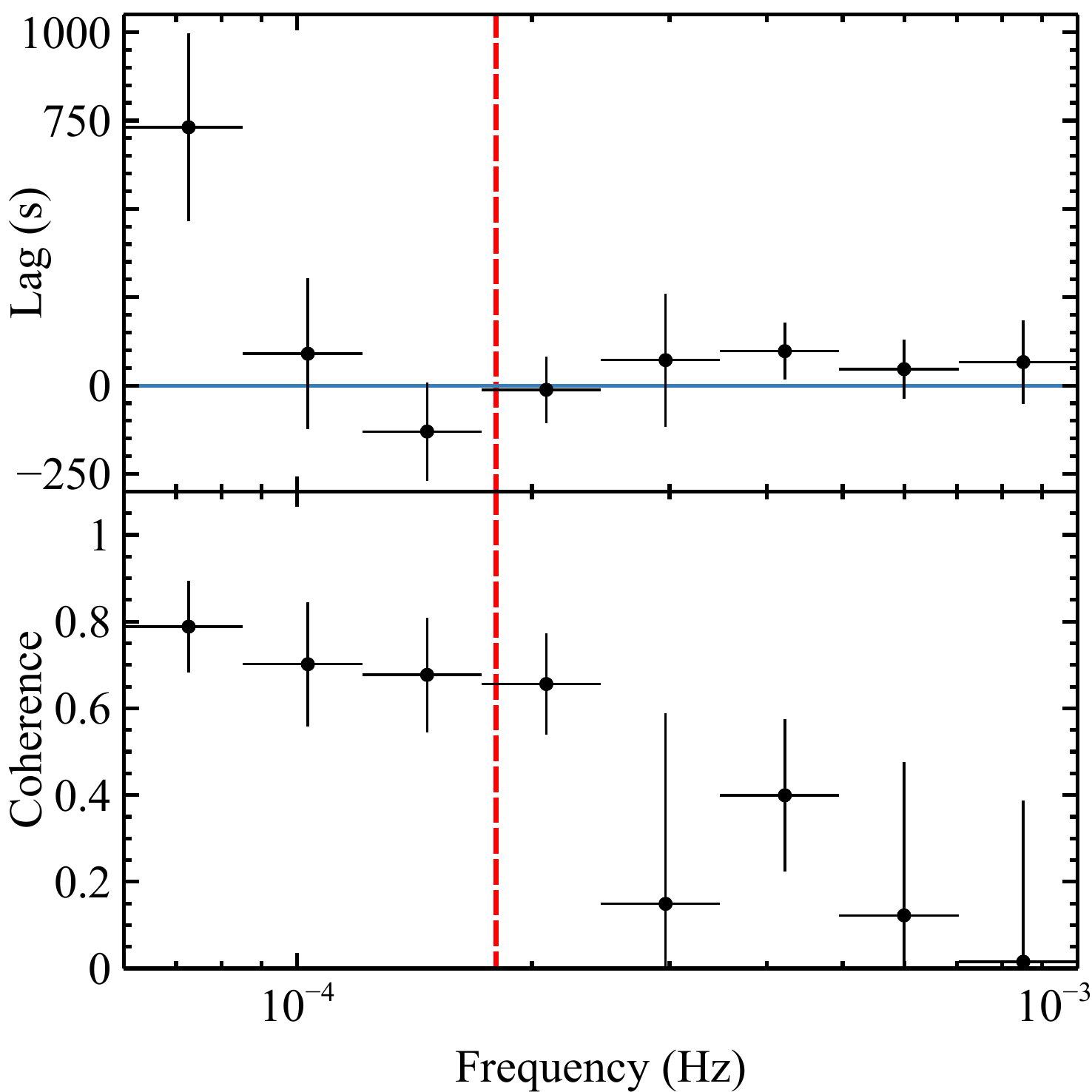}
    \caption{Time lag and coherence between variations in the soft (0.1-3 keV) and hard (1-4 keV) bands as a function of timescale (temporal frequency). The dashed red line is the frequency above which we would have expected reverberation. Our results yield a soft lag of zero, possibly suggesting the soft excess is not necessarily due to reflection, and a relatively low coherence, suggesting that the soft excess may not be directly correlated with the hard band.}
    \label{fig:coh}
\end{figure}

Figure \ref{fig:coh} shows the coherence-frequency spectrum that results from our data. In the frequency range where we expect to see reverberation, given this black hole mass, we can see that for the most part the coherence is below 0.6 and consistent with 0. The non-unity coherence at these frequencies shows that there is a non-linearly correlated component in the soft band. A coherence less than unity may have to do with an additional soft excess continuum component that is variable in a way that is not correlated with the power-law and reflection components.

\section{Discussion}
\label{sec:disc}
\subsection{Where does the X-ray emission of 1H 0323+342 come from?}

Previous studies of 1H 0323+342 have had mixed approaches when it comes to the modeling of the source's X-ray spectra. By using a simple comparison in count rates between the soft (0.3-2 keV) and hard (2-10 keV) bands of \textit{Swift}-XRT observations, \cite{2014ApJ...789..143P} showed that 1H 0323+342 might exhibit a strong soft excess whose variability was not perfectly correlated with the variability of the hard band, which may suggest that at least two spectral components are required to fit the X-ray spectra of this source. They successfully fit the spectrum using an ionized reflection model, and found that the fit improved when they took relativistic blurring into account. However, they fix the inclination to $10$ deg. This resulted in a steep spectrum ($\Gamma = 2.02\pm0.06$) and a high spin ($a = 0.96\pm0.14$). \citet{2018MNRAS.479.2464G} also assumed that the soft excess was due to reflection and arrived at similar results. However, other studies such as \citet{2019ApJ...872..169P} and \cite{2018MNRAS.475..404K} have managed to obtain good fits by modeling the soft excess with a power law, which adds a layer of ambiguity regarding the best description of the excess at low energies. 

In each of these cases (as well as in \citealt{2013MNRAS.428.2901W}, \citealt{2015AJ....150...23Y}), there was some evidence of a potential broad Fe K$\alpha$ line, but there were a few aspects that were unclear, such as a stark difference in the measured black hole spin between \citet{2013MNRAS.428.2901W} and \citet{2015AJ....150...23Y} (although the latter froze their inclination to a much lower value), and the fact that the emission was relatively weak to begin with. In addition, \citet{2019ApJ...872..169P} and \citet{2018MNRAS.475..404K} found that a combination of narrow emission lines was adequate enough to fit the residuals at $\sim6$ keV. Therefore, as for the soft excess, the nature of these residuals was up for debate.

We find through our spectral analysis that reflection may play a role in both the excess emission in the soft band and the residuals at higher energies ($\sim$2-35 keV); that being said, regardless of the model we use, we require some additional soft excess component that is not reflection. A combination of a reflection model (\texttt{relxill}) and a phenomenological blackbody model, along with a distant reflector, results in a good fit to most of the broadband spectrum, and we detect an iron line at $6.6\pm0.1$ keV thanks to the increased signal-to-noise ratio that results from combining 6 simultaneous \textit{XMM-Newton} and \textit{NuSTAR} observations with archival data.  However, a discussion of the origin of the blackbody-like component is warranted, and whether it is related to the jet or is due to Comptonization is still unclear. 

A high density reflection model can also describe the data fairly well, but certain well-known issues arise, such as the fact that the featureless soft excess requires stronger broadening than the fit to the iron line alone. An alternative ionized partial covering model also works; however, the variability imprinted by a partial covering model would not predict the linear RMS-flux relation we obtain, which suggests that the variability of this source is multiplicative in nature. This further suggests that the X-ray emission originates in the central region of the AGN.

While we detect a hard low-frequency lag that suggests propagating disk fluctuations, we do not detect a soft lag. RQ AGN with similar variance, exposure, and flux do show soft lags, which might suggest that this is not a signal-to-noise issue. This might have to do with the presence of an additional soft excess component that is not simply blurred reflection and that displays variability that is uncorrelated with the continuum-dominated hard band.

The final piece of the spectral puzzle involves the possible hard excess observed at energies above $\sim35$ keV. We find that, in each of our models, the residuals seem to increase at higher energies. None of our reflection models were able to take care of this, which implies that this excess may be due to non-thermal processes related to the jet. The otherwise coronal-dominated X-ray spectrum suggests the emission might be contaminated by a combination of synchrotron self-Compton mechanisms and Comptonization of external disk photons. This hard excess was also observed in \citet{2018MNRAS.479.2464G}. \citet{2014ApJ...789..143P} and \citet{2019ApJ...872..169P} also detect a hard excess and manage to fit it with a broken power law and a second power law, respectively, and they attribute this to a combination of thermal coronal emission and jet emission with a power-law shape, the latter of which would result from the AGN undergoing $\gamma$-ray flaring. In other words, the same process that produces the $\gamma$-ray emission (such as synchrotron self-Compton or external Compton) may be the dominant form of hard X-ray emission above 35 keV. The fact that an additional hard power law component helps fit the hard excess we observe at a significance of $>99.99\%$ seems to suggest a similar behavior.

\subsection{Model Inconsistencies and Alternative Assumptions}
\citet{2016RAA....16..176F} constrain the viewing angle of 1H 0323+342 to values between 4 and 13 degrees using observations in the radio band. In X-ray spectra, the inclination can be independently determined by the shape of the Fe K$\alpha$ line in the reflection spectrum: if the line is mostly broad and blue-shifted, line of sight Doppler boosting (i.e. larger line-of-sight velocities) has the dominant influence on the line profile, and therefore this would correspond to a highly inclined disk; on the other hand, if the profile is mostly skewed towards lower energies, it would mean that gravitational and time dilation redshifts are dominant, corresponding to a disk that is almost face-on.

Our results in both the 2-79 keV and the broadband ranges with \texttt{relxill} provide an inclination that is higher than the upper limit obtained from radio observations (see Table \ref{tab:table3}). Our detection of the iron line also shows that the line is blue-shifted. \citet{2013MNRAS.428.2901W} had modeled \textit{Suzaku} observations with relativistic reflection and found that if they let the inclination vary, they end up with a highly blue-shifted line that peaks at $\sim8$ keV, and obtain an unphysical value of $i=82\pm3$ deg. Therefore, high inclinations have been suggested in the past; however, as previously mentioned, the iron line had only been marginally detected. In this work, we finally have a constraint on the iron line profile, and we still measure a high inclination. Given the latter, it seems that the broadening of the line can be explained almost exclusively through line-of-sight Doppler effects, but this is a lot more Doppler boosting than is physically allowed by the inclination measured in the radio observations. This, along with the unusually flat emissivity profile from our best fit, seems to suggest that we may have to reconsider our model assumptions.

One way to account for a perceived high inclination is to consider the possibility that the emission is being viewed through the acceleration zone of the jet. In this scenario, a highly energetic corona would end up inverse Compton scattering the reflection spectrum. Studies have shown that considering these processes provides a more complete picture of the emission from both black hole transients and AGN, with the logic being that if thermal seed photons from the disk are upscattered by the corona, reprocessed reflection photons that emerge in the inner regions of the disk should also be influenced by the same process \citep[e.g.,][]{2015MNRAS.448..703W,2017ApJ...836..119S}. This process could cause the iron line in 1H 0323+342 to be blueshifted, even if the disk is face-on and the jet is closely aligned to the line of sight.

Another way to explain our measurements may involve abandoning the thin disk geometry \citep{1973A&A....24..337S} altogether and considering a model where the accretion disk is geometrically thick, a scenario which has also been suggested to explain similar inclination mismatches in X-ray binaries \citep[e.g.,][]{2019ApJ...882..179C}. In the past few years, simulations have been used to understand the underlying physical processes behind sources that have very high accretion rates \citep[e.g.,][]{2015MNRAS.454L...6M,2015MNRAS.453.3213S}. All of these simulations lead to an accretion disk that is both optically and geometrically thick. This highly accreting, geometrically thick scenario could lead to optically thick outflows that restrict the emitted radiation to a funnel-like space. A jet from near the black hole could then be able to clear out some of the inner flow and could potentially give us a line of sight towards the central X-ray emitting region \citep{2015MNRAS.454L...6M}. 

If 1H 0323+342 were to have a relatively high accretion rate, its accretion disk would not be infinitesimally thin. Since radio observations point to a low-inclination geometry, the blue-shifted iron line we observe could be caused by radiation-driven outflows in the funnel, and since the presence of the jet would be exposing the central region, gravitational redshifts could also be used to explain the broadening of the line. \citet{2016Natur.535..388K} develop a simple model for iron line reverberation in a funnel geometry where special relativistic effects and gravitational redshift are taken into account, and where only an outflow velocity profile for the funnel walls is considered. Through Monte Carlo simulations, they show that an iron line can indeed be blueshifted and broadened due to reflection off an optically thick outflow.

A very recent study by \citet{2019ApJ...884L..21T} makes use of this funnel geometry to make a detailed comparison between the iron line profiles of thin disks and those of a geometrically thick supper-Eddington disk. In particular, they show that, in each type of disk, photons emitted from different radii have different contributions to the line profile. They show that for thin disks, most of the flux comes from radii less than $25R_{\rm g}$, since the emissivity profile drops relatively quickly at larger radii. However, for a funnel geometry, more photons are emitted between $25-50R_{\rm g}$ because the reflected flux does not decrease as quickly at large radii, due to the fact that in this geometry, the emission from the corona would be able to illuminate more of the disk at larger radii than it would for a thin disk. In addition, they show that the iron line profiles of the thick disk are significantly more blue-shifted than those of the thin disk, for reasons mentioned earlier.

We can test if 1H 0323+342 would be an adequate candidate for a thick-disk model by performing a quick calculation to estimate its Eddington rate using our data. We use the 2-10 keV luminosity from our best fit model to estimate the latter. After performing this calculation, we arrive at a value of log($L_{\rm 2-10 \ keV}$)=43.8. Adjusting the appropriate value in ergs s$^{-1}$ by a bolometric correction of $\sim$20 \citep{2007MNRAS.381.1235V} to get the bolometric luminosity $L = \eta \dot{M}c^{2}$, we then divide by the Eddington luminosity.
As in our timing analysis, we use the mass $M_{\rm BH} = 1.9^{+0.4}_{-0.3}\times10^{7}$M\textsubscript{\(\odot\)} obtained by \citet{2017MNRAS.464.2565L}. Our result is then $L \sim 0.5^{+0.2}_{-0.1} L_{\rm Edd}$. The values we calculate here are comparable to the luminosity and Eddington ratio values obtained in \citet{2018MNRAS.475..404K} and several other previous works \citep[e.g.,][]{2019ApJ...872..169P,2014ApJ...789..143P,2018ApJ...866...69P,2017MNRAS.464.2565L}. Although this ratio does not imply super-Eddington accretion, it does represent a relatively high accretion rate.

It is important to note that the bolometric correction we use is really an average obtained for Eddington ratios $\lesssim$0.1. \citet{2007MNRAS.381.1235V} show that the correction seems to increase with Eddington ratio, and discuss that it may vary considerably between different objects, in particular when it comes to sources like NLS1s and RL AGN. In fact, according to \citet{2007MNRAS.381.1235V}, for high-Eddington AGN, the bolometric correction is closer to 60. Given the high accretion rate of NLS1s, this could mean that 1H 0323+342 may have an Eddington ratio as high as $\sim1.5$. Our results seem to suggest that these are the physical phenomena that are relevant for the case of 1H 0323+342.

\section{Conclusions}

We have presented X-ray spectral and timing analyses of the $\gamma$-NLS1 1H 0323+342 with a combination of the first simultaneous \textit{XMM-Newton} and \textit{NuSTAR} campaign and archival data. Our main results are as follows:

\begin{enumerate}
    \item We definitively measure a broad iron line and a Compton hump in 1H 0323+342. The broad line extends from $\sim5.5-8$ keV, meaning that there are significantly blueshifted and redshifted components of the line.
    \item We attempt describing the data with a variety of absorption and reflection based models. We find that the X-ray emission in the 0.5-79 keV range (soft excess and iron line) is well described by a combination of a phenomenological blackbody component and relativistic reflection model, suggesting that the dominant form of the X-ray emission comes from the corona in the vicinity of the black hole.
    \item We find potential evidence of a hard excess at high energies, suggesting that non-thermal jet emission contributes yet another component to the hard X-ray spectrum.
    \item Our measurement of the inclination using a razor-thin disk model is definitively in tension with radio observations at a significance of >99.99\%; this measurement can be explained by the detection of a blue-shifted line in the spectrum. However, the discrepancy might be reconciled by considering a model that takes into account a geometrically thick accretion flow.
\end{enumerate}

\section*{Acknowledgements}
SM and EK acknowledge support from NASA grant 80NSSC20K1085.
%%%%%%%%%%%%%%%%%%%%%%%%%%%%%%%%%%%%%%%%%%%%%%%%%%



\bibliographystyle{mnras}
\bibliography{refernces} % if your bibtex file is called example.bib

%%%%%%%%%%%%%%%%%%%%%%%%%%%%%%%%%%%%%%%%%%%%%%%%%%

%%%%%%%%%%%%%%%%% APPENDICES %%%%%%%%%%%%%%%%%%%%%

\appendix

\section{Checking for a potential absorption feature between 8 and 10 \texorpdfstring{\MakeLowercase{ke}V}{keV}}
\label{appendix:A}

\begin{figure}
	    \centering
	    \includegraphics[width=\columnwidth]{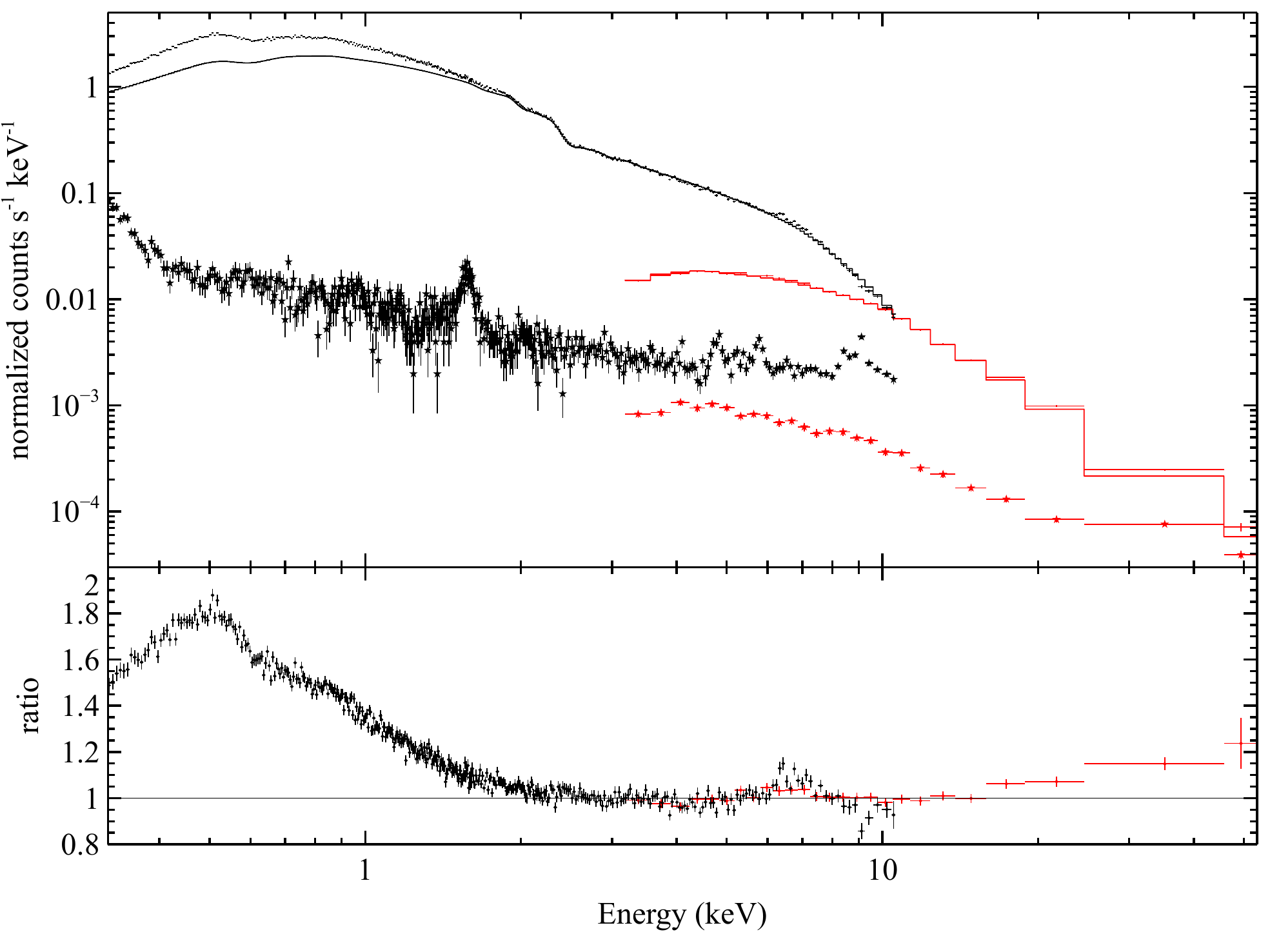}
        \caption{2-79 keV fit with a power law, extrapolated to lower  energies. The corresponding background spectrum is also shown, with a peak at the same energies as our negative residuals.}
        \label{fig:softex}
    \end{figure}

The background spectrum reveals a peak that corresponds to the energies where we observe negative residuals (Figure \ref{fig:softex}). This is likely the Cu-K$\alpha$ complex in EPIC-pn, and our negative residuals are therefore the result of over-subtraction of these features. To be more thorough, we perform an exercise to make sure that the negative residuals are not the result of a physical absorption process. We fit the EPIC-pn spectrum, without the background subtracted, to a power law; in this case, if there is indeed absorption from an outflow, for instance, the residuals should remain negative, indicating that the feature is not just an artifact.

\begin{figure}
	    \includegraphics[width=\columnwidth]{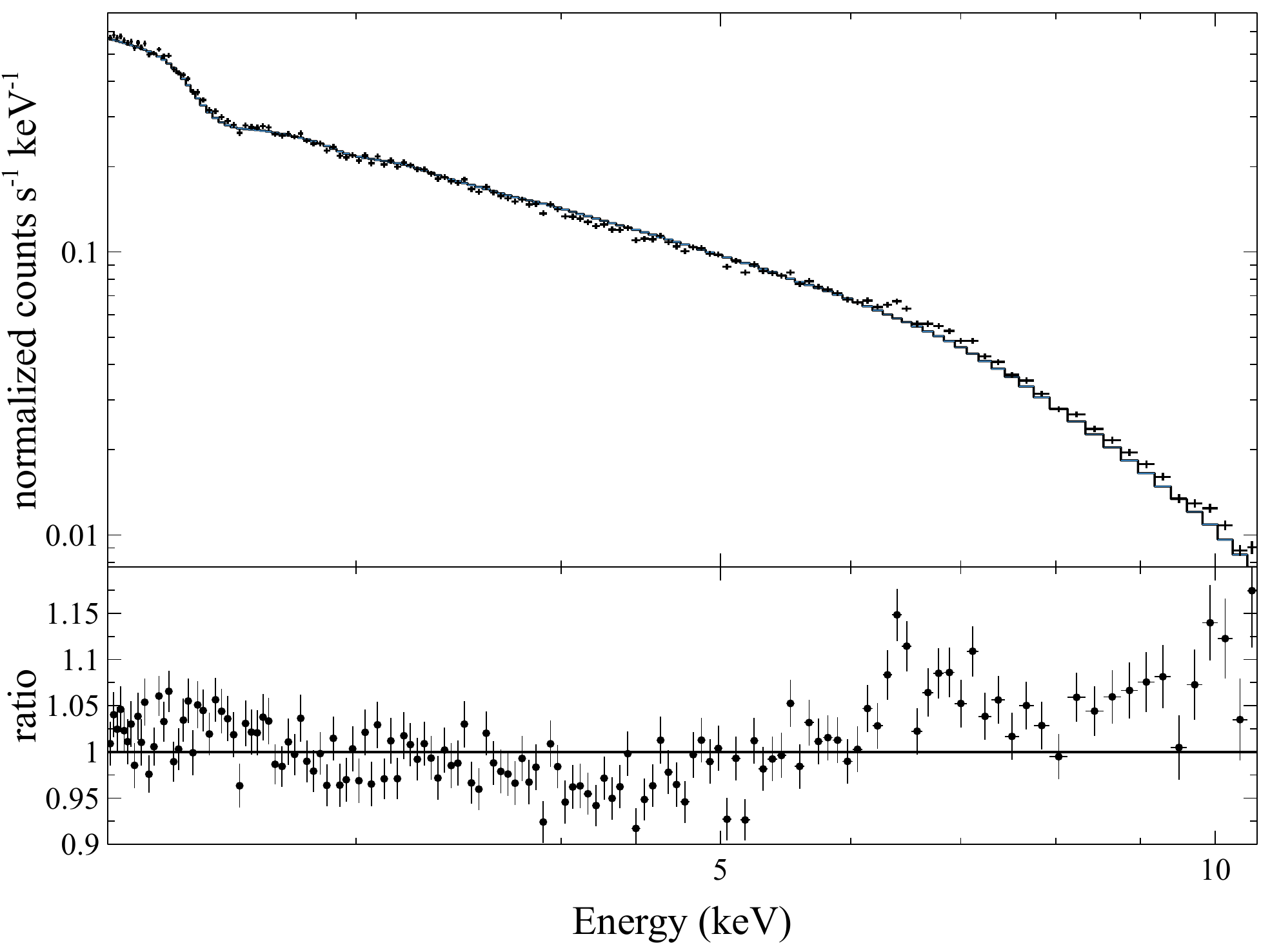}
        \caption{2-10 keV spectrum, with no background subtraction, fit to a power law. The negative residuals we observe in our main analysis are not present, suggesting that they are an artifact caused by background over-subtraction.}
        \label{fig:nobacksub}
    \end{figure}

We show this spectrum in Figure \ref{fig:nobacksub}, and we find that the residuals are not present, suggesting that background over-subtraction is at fault. We also perform this exercise for our individual high and low-flux observations, and still the negative residuals do not show up. In addition, none of the individual MOS spectra show these residuals.

%%%%%%%%%%%%%%%%%%%%%%%%%%%%%%%%%%%%%%%%%%%%%%%%%%

% Don't change these lines
\bsp	% typesetting comment
\label{lastpage}
\end{document}